\documentstyle[aps,prl,twocolumn,psfig]{revtex}
\begin{document}\hbadness=10000
\twocolumn[\hsize\textwidth\columnwidth\hsize\csname
@twocolumnfalse\endcsname
\title{Expected Production of Strange Baryons and 
Antibaryons in Baryon-Poor {\bf QGP}}
\author{Johann Rafelski     and Jean Letessier}

\address{ 
        Department of Physics,
        University of Arizona,
        Tucson, AZ 85721, USA\\
        LPTHE,Universit\'e Paris 7,
        2 place Jussieu, F--75251 Cedex 05
}
\date{August 5, 1999}
\maketitle
\begin{abstract}\noindent 
{In a dynamical model of {\small QGP} at {\small RHIC}
we obtain the temporal evolution of
strange phase space occupancy at conditions 
expected to occur in 100+100$A$\,{\small GeV}
nuclear collisions. We show that the sudden 
{\small QGP} break up model developed to 
describe the {\small SPS} experimental results implies
dominance of both baryon and antibaryon abundances 
by the strange baryon and antibaryon 
yields.\vskip 0.3cm

PACS number(s):  12.38.Mh, 25.75.-q
}
\end{abstract}\vskip1.5pc]
%%%%%%%%%%%%%%%%%%%%%%%%%%%%%%%%%%%%%%%%%%%%%%%%%%%%%
We explore the consequences of high strangeness 
abundance we expect to be present in the baryon-poor 
quark-gluon plasma ({\small QGP}) environment formed
{\it e.g.}, in the central rapidity region 
in Au--Au, maximum energy 100+100$A$ {\small GeV}
collisions at the relativistic heavy ion 
collider ({\small RHIC}) at the Brookhaven National Laboratory, 
Upton, New York. About 10--20\% of hadrons 
produced  in these reactions will be strange, 
and since mesons dominate  hadron  
abundance, there is much more strangeness than baryon number. 
During the  break-up of the color charge deconfined {\small QGP} 
phase there is considerable
advantage for strangeness to stick to baryons given that 
the  energy balance for the same flavor content 
favors production of strange baryons 
over kaons, ($E(\Lambda+\pi)<E(\mbox{N+K})$)\,.  
When {\small QGP}  is formed, we therefore expect 
to find hyperon dominance of baryon distribution
{\it i.e.} most baryons and antibaryons produced at {\small RHIC} 
will be strange. A similar argument could be made for reactions 
leading to the confined phase, the main
difference arises from the observation that the required 
high abundance of strangeness per participating nucleon 
can be produced in the deconfined 
{\small QGP} \cite{RM82,MS86,BCD95,acta96}. 
This qualitative argument will be quantitatively elaborated here,
in view of the  considerable effort that has
been committed by the {\small STAR} collaboration at {\small RHIC}
to enhance the capability to measure 
multi-strange (anti)baryon production 
using a silicon strip detector (SSD) \cite{stars}. 
 
In the first part of this report we show that the chemical strangeness
flavor abundance equilibrium is established at the time of 
{\small QGP} breakup by the dominant process 
which is  gluon fusion, $GG\to s\bar s$. In {\small QGP} this
reaction overcomes the current quark mass threshold $2m_{\rm s}$ 
for strangeness formation for temperature 
$T\simeq m_{\rm s}> 200$\,{\small MeV}. In the second part of this
report we use the computed chemical condition of strangeness, 
and employ the knowledge gained in our 
analysis of the {\small SPS}  results \cite{recent}, to obtain the 
strange baryon and antibaryon abundances
expected at {\small RHIC}. 

In some key aspects the methods  we employ
differ from those obtained in other studies of chemical 
equilibration of quark flavor for {\small RHIC} 
conditions \cite{Bir93,Won96,Sri97}. We study 
the dynamics of the phase space
occupancy rather than particle density,
 and we eliminate most of the 
dynamical flow effects by considering  entropy 
conserving  evolution. Moreover,  we
use  running {\small QCD} parameters (both 
coupling and strange quark mass) to describe 
strangeness production, with strong 
coupling constant $\alpha_s$  as determined 
 at the $M_{Z^0}$ energy scale.
We will make two assumptions of relevance for the
results we obtain:\\
\noindent a) the kinetic (momentum distribution) 
equilibrium is reached faster than the chemical (abundance) 
equilibrium \cite{Shu92,Alam94};\\
\noindent b) gluons  equilibrate chemically significantly 
faster than strangeness \cite{Won97}.\\
The first assumption  allows to 
study only the chemical abundances, rather than
the full momentum distribution, which simplifies greatly
the structure of the master equations; the second  assumption 
allows to focus after an initial time $\tau_0$ has passed 
on the evolution of strangeness population: $\tau_0$ 
is the time required for the development to near chemical
equilibrium of the  gluon population.

We now formulate the dynamical equation for the 
evolution of the phase space occupancy $\gamma_{\rm s}$ 
of strange quarks in the expanding {\small QGP}: 
the phase space  distribution  $f_{\rm s}$  
can be  characterized by a local 
temperature $T(\vec x,t)$ of a (Boltzmann) equilibrium distribution  
$f_{\rm s}^\infty$\,, with  normalization set 
by a phase space occupancy factor:
\begin{equation}\label{gamdef2}
f_{\rm s}(\vec p,\vec x; t))\simeq \gamma_{\rm s}(T) 
   f_{\rm s}^\infty(\vec p;T)\,.
\end{equation}
Eq.\,(\ref{gamdef2})  invokes in the momentum independence of 
$\gamma_{\rm s}$ our  first assumption. More generally,
the factor $\gamma_i,\, i=g,q,s,c$ allows a local density 
of gluons, light quarks, strange quarks and charmed quarks,
respectively not to be 
determined by the local momentum shape, but to evolve 
independently.  

With variables 
$(t,\vec x)$ referring to an observer in the laboratory 
frame, the chemical evolution can be described by the strange
quark  current non-conservation arising from strange quark 
pair production described by a Boltzmann collision term:
\begin{eqnarray}
\partial_\mu j^\mu_{\rm s}& & \equiv{\partial \rho_{\rm s}\over \partial t} +
   \frac{ \partial \vec v \rho_{\rm s}}{ \partial \vec x}= \frac12
\rho_g^2(t)\,\langle\sigma v\rangle_T^{gg\to s\bar s}+\nonumber\\ & & 
 +\;\rho_q(t)\rho_{\bar q}(t)
\langle\sigma v\rangle_T^{q\bar q\to s\bar s}\! -
\rho_{\rm s}(t)\,\rho_{\bar{\rm s}}(t)\,
\langle\sigma v\rangle_T^{s\bar s\to gg,q\bar q}.
\label{drho/dt1}
\end{eqnarray}
The factor 1/2  avoids double counting of gluon pairs.
The implicit sums over spin, color and any other 
discreet quantum numbers are combined in the particle density 
 $\rho=\sum_{s,c,\ldots}\int d^3p\,f$, and  we have also 
introduced the  momentum averaged production/annihilation
thermal reactivities:
\begin{equation}
\langle\sigma v_{\rm rel}\rangle_T\equiv
\frac{\int d^3p_1\int d^3p_2 \sigma_{12} v_{12}f(\vec p_1,T)f(\vec p_2,T)}
{\int d^3p_1\int d^3p_2 f(\vec p_1,T)f(\vec p_2,T)}\,.
\end{equation}
$f(\vec p_i,T)$ are the relativistic Boltzmann/J\"uttner
distributions of two colliding particles $i=1,2$ of momentum $p_i$. 

The  current conservation  can also be written with reference to the 
individual particle dynamics \cite{LambH}: consider
$\rho_{\rm s}$  as the  inverse of the small
volume available to each particle. Such a volume is defined 
in the local frame of reference for which 
the local flow vector vanishes $\vec v(\vec x,t)|_{\mbox{local}}=0$.
The considered volume $\delta V_l$ being 
occupied by small number of particles 
$\delta N$ ({\it e.g.}, $\delta N=1$), we have:
  \begin{equation}\label{Nsinf}
\delta N_{\rm s}\equiv \rho_{\rm s} \delta V_l \,.
%=\gamma_{\rm s}\delta V_l \rho_{\rm s}^\infty
\end{equation}
 The left hand side ({\small LHS}) 
of Eq.\,(\ref{drho/dt1}) can be now written as: 
\begin{equation}\label{LagCor}
{\partial \rho_{\rm s}\over \partial t} +
   \frac{ \partial \vec v \rho_{\rm s}}{ \partial \vec x}\equiv
       \frac{1}{\delta V_l}\frac{d\delta N_{\rm s}}{dt}= 
\frac{d\rho_{\rm s}}{dt}+\rho_{\rm s}\frac1{\delta V_l}\frac{d\delta V_l}{dt}\,.
\end{equation}
Since $\delta N$ and $\delta V_l dt$ are L(orentz)-invariant, 
the actual choice of the frame of reference in which the 
right hand side ({\small RHS}) of Eq.\,(\ref{LagCor}) is 
studied is irrelevant and we drop henceforth the subscript $l$.

We can further adapt Eq.\,(\ref{LagCor}) to the dynamics we pursue:
we introduce  
$\rho_{\rm s}^\infty(T)$ as the (local) chemical equilibrium  abundance 
of strange quarks, thus $\rho=\gamma_{\rm s}\rho_{\rm s}^\infty$.
We evaluate the equilibrium abundance  
$\delta N_{\rm s}^\infty=\delta V\rho_{\rm s}^\infty(T)$
 integrating the Boltzmann distribution:
\begin{equation}\label{Nsinfty}
\delta N_{\rm s}^\infty=[\delta VT^3] {3\over\pi^2} \,z^2K_2(z)\,,
\quad z={m_{\rm s}\over T}\,,
\end{equation}
where $K_\nu$ is the modified Bessel function of order $\nu$; we 
will below use: $d[z^\nu K_\nu(z)]/dz=-z^\nu K_{\nu-1}$\,.
The first factor on the {\small RHS}
 in Eq.\,(\ref{Nsinfty}) is a constant in time should the
evolution of matter after the
initial pre-thermal time period $\tau_0$ 
be entropy conserving \cite{Bjo83}, and thus 
 $\delta VT^3=\delta V_0T^3_0$=Const.\,.
We now substitute in Eq.\,(\ref{LagCor}) 
and obtain
\begin{equation}\label{lochem}
{\partial \rho_{\rm s}\over \partial t} +
   \frac{ \partial \vec v \rho_{\rm s}}{ \partial \vec x}=  
\dot T\rho_{\rm s}^\infty\left({{d\gamma_{\rm s}}\over{dT}}+
\frac{\gamma_{\rm s}}{T}z\frac{K_1(z)}{K_2(z)}\right)\,,
\end{equation}
where $\dot T=dT/dt$. Note that in Eq.\,(\ref{lochem}) 
only  a part of  the usual flow-dilution term
is left, since we implemented the adiabatic volume expansion,
and study the evolution of
the phase space occupancy in lieu of particle density.
The dynamics of the local temperature is the only quantity we need
to model.

We now return to study the collision terms seen on
the {\small RHS} of Eq.\,(\ref{drho/dt1}). 
A related  quantity is the (L-invariant)
production  rate $A^{12\to 34}$ of particles 
per unit time and space, defined usually 
with respect to chemically equilibrated distributions: 
\begin{equation}\label{prodgen}
A^{12\to 34}\equiv\frac1{1+\delta_{1,2}} \rho_1^\infty\rho_2^\infty 
             \langle \sigma_{\rm s} v_{12}\rangle_T^{12\to 34}  \,.
\end{equation}
The factor $1/(1+\delta_{1,2})$ is introduced to compensate 
double-counting of identical particle pairs. 
In terms of the L-invariant $A$\,,
% Eq.\,(\ref{drho/dt1}) and the 
% flow term seen in the local reference frame, Eq.\,(\ref{lochem}),
Eq.\,(\ref{drho/dt1}) takes the form:
\begin{eqnarray}
&&\dot T\rho_{\rm s}^\infty\left({{d\gamma_{\rm s}}\over{dT}}+
\frac{\gamma_{\rm s}}{T}z\frac{K_1(z)}{K_2(z)}\right)
=
\gamma_g^2(\tau)A^{gg\to s\bar s} +\nonumber\\
&&+\gamma_q(\tau)\gamma_{\bar q}(\tau)A^{q\bar q\to s\bar s} 
\!-\gamma_{\rm s}(\tau)\gamma_{\bar s}(\tau)(A^{s\bar s\to gg}
\!+A^{s\bar s\to q\bar q}).
\label{rho-gam}
\end{eqnarray}
Only weak interactions convert quark flavors, thus, 
on  hadronic time scale, we have 
$\gamma_{s,q}(\tau)=\gamma_{{\bar s},{\bar q}}(\tau)$. Moreover, 
detailed balance, arising from the time reversal symmetry of the
microscopic reactions, assures that the invariant rates
for forward/backward reactions are the same, specifically
\begin{equation}\label{det-bal}
A^{12\to 34}=A^{34\to 12},
\end{equation}
and thus:
\begin{eqnarray}
\dot T\rho_{\rm s}^\infty\left({{d\gamma_{\rm s}}\over{dT}}+
\frac{\gamma_{\rm s}}{T}z\frac{K_1(z)}{K_2(z)}\right)
=&&
\gamma_g^2(\tau)A^{gg\to s\bar s}
    \left[1-\frac{\gamma_{\rm s}^2(\tau)}{\gamma_g^2(\tau)}\right] \nonumber\\
&&\hspace{-1cM}
+\gamma_q^2(\tau)A^{q\bar q\to s\bar s}
    \left[1-\frac{\gamma_{\rm s}^2(\tau)}{\gamma_q^2(\tau)}\right]\,.
\label{rho-bal}
\end{eqnarray}
When all $\gamma_i\to 1$, the Boltzmann collision term vanishes, we have
reached equilibrium.

As discussed, the gluon chemical equilibrium
is thought to be reached at high temperatures well before the strangeness
equilibrates chemically, and thus we assume this in what follows, and 
the initial conditions we will study refer to the time at which gluons
are chemically equilibrated. Setting $\lambda_g=1$ (and without a
significant further consequence for what follows, 
since gluons dominate the production rate, also $\lambda_q=1$)
we obtain after a straightforward manipulation the dynamical 
equation describing the evolution of the local phase space occupancy
of strangeness:
\begin{equation}\label{dgdtf}
2\tau_{\rm s}\dot T\left({{d\gamma_{\rm s}}\over{dT}}+
\frac{\gamma_{\rm s}}{T}z\frac{K_1(z)}{K_2(z)}\right)
=1-\gamma_{\rm s}^2\,.
\end{equation}
Here, we  defined the relaxation time  $\tau_{\rm s}$  of 
chemical (strangeness) equilibration as the 
ratio of the equilibrium density that
is being approached, with the rate at which this occurs:
\begin{equation}\label{tauss}
\tau_{\rm s}\equiv
{1\over 2}{\rho_{\rm s}^\infty\over{
(A^{gg\to s\bar s}+A^{q\bar q\to s\bar s}+\ldots)}}\,.
\end{equation}
The factor 1/2 is introduced by convention in order for the quantity
$\tau_{\rm s}$ to describe the exponential approach to equilibrium.

Eq.\,(\ref{dgdtf})  is our final analytical result describing 
the  evolution of phase space  occupancy.
Since one generally  expects that $\gamma_s\to 1$ in a monotonic 
fashion as function of time, it is important to appreciate that this 
equation allows  $\gamma_{\rm s}>1$:
when $T$ drops below  $m_{\rm s}$, and $1/\tau_{\rm s}$ becomes small,
the dilution term (2nd term on {\small LHS})
in Eq.\,(\ref{dgdtf})  dominates the evolution of $\gamma_{\rm s}$\,.
In simple terms,  the high
abundance of strangeness produced at high temperature over-populates
the available phase space at lower temperature, when the equilibration
rate cannot keep up with the expansion cooling.
This behavior of $\gamma_{\rm s}$  has been 
shown in Fig.\,2 of Ref\cite{Let97} for the {\small SPS} 
conditions with fast transverse expansion. 
Since we assume that the dynamics of transverse expansion 
of {\small QGP} is similar at {\small RHIC} as at {\small SPS}, 
we will obtain a rather similar behavior for $\gamma_{\rm s}$. 
We note that yet a faster transverse  expansion than considered here 
could enhance the chemical strangeness 
anomaly.

$\tau_{\rm s}(T)$\,, Eq.\,(\ref{tauss}),
has been evaluated using p{\small QCD} cross section and
 employing {\small NLO} (next to leading order)
running of both the strange quark mass and {\small QCD}-coupling 
constant $\alpha_s$ \cite{budap}. We believe that this method produces
a result for $\alpha_s$ that can be trusted  
down to 1\,{\small GeV} energy scale which
is here relevant. We employ results obtained with
$\alpha_{s}(M_{Z^0})=0.118$ and 
$m_{\rm s}(\mbox{1{\small GeV}})=220$\,{\small MeV}, a
somewhat conservative (high) choice for $m_{\rm s}$, 
which should under-predict strangeness production.
There is some systematic  uncertainty due to the appearance of the 
strange quark mass as a fixed rather than running value
in both, the chemical equilibrium 
density $\rho_{\rm s}^\infty$ in Eq.\,(\ref{tauss}), and in the dilution term
in Eq.\,(\ref{dgdtf}). We use the value 
$m_{\rm s}(\mbox{1\,{\small GeV}})$, with the energy
scale chosen to correspond to typical interaction scale in the QGP. 

Numerical study of Eq.\,(\ref{dgdtf}) becomes possible  
as soon as we define the temporal evolution of the 
temperature for {\small RHIC} conditions.   
We expect that a global cylindrical expansion 
should describe the dynamics: aside of the longitudinal flow, we 
allow  the cylinder surface to expand given the internal thermal pressure.
{\small SPS} experience suggests that the transverse matter flow will not 
exceed the sound velocity of relativistic matter $v_\bot\simeq c/\sqrt{3}$.
We recall that for pure longitudinal expansion local entropy density scales
as $S\propto T^3\propto 1/\tau$, \cite{Bjo83}. 
It is likely that the transverse flow of matter 
 will accelerate the drop in entropy density. We thus
 consider the following  temporal evolution function 
of the temperature:
\begin{equation}\label{Toft}
T(\tau)=T_0\left[
\frac{1}{(1+\tau\ 2c/d)(1+\tau\ v_\bot/R_\bot)^2}
\right]^{1/3}\,.
\end{equation}
We take the thickness of the initial collision region at $T_0=0.5$\,{\small GeV}
to be\  $d(T_0=0.5)/2=0.75$\,fm,  and the transverse dimension in nearly 
central Au--Au  collisions to be $R_\bot=4.5$\,fm. 
The time at which thermal initial conditions are reached is
assumed to be $\tau_0=1$fm/$c$. When we vary $T_0$, 
the temperature  at which the  gluon equilibrium is reached, we also
scale the longitudinal dimension according to:
\begin{equation}\label{dofT0}
d(T_0)=(0.5\mbox{\,{\small GeV}}/T_0)^3 1.5\mbox{\,fm}\,.
\end{equation}
This assures that when comparing the different evolutions of $\gamma_s$ we
are looking at an initial system that has the same entropy 
content. The reason we vary the initial temperature $T_0$
down to 300 {\small MeV}, maintaining the initial entropy content
is to understand how the assumption about the 
chemical equilibrium of gluons, reached by definition at  $T_0$,
impacts our result.

The numerical integration of Eq.\,(\ref{dgdtf}) is started at 
$\tau_0$, and a range of initial temperatures $300\le T_0\le 600$, 
varying in steps of 50 {\small MeV}.   The high limit of the temperature
we  explore exceeds somewhat the ``hot glue scenario'' \cite{Shu92},
while the lower limit of $T_0$ corresponds to the more conservative
estimates of possible initial conditions \cite{Bjo83}.
Since the initial $p$--$p$ collisions also
produce strangeness,  we take as an estimate of initial 
abundance a common initial value
$\gamma_{\rm s}(T_0)=0.2$. The time evolution in 
the plasma phase is followed up to
the break-up of QGP.  This condition we establish
in view of our analysis of the {\small SPS} results. We recall
that {\small SPS}-analysis showed that the system dependent 
baryon and antibaryon  $m_\bot$-slopes of particle spectra are
result of differences in collective  flow in the deconfined 
{\small QGP} source at freeze-out \cite{recent}. There is a universality of
physical properties of hadron chemical  freeze-out between different
{\small SPS} systems, and  in our analysis a practical coincidence of 
the kinetic  freeze-out conditions with the chemical freeze-out.
 We thus expect extrapolating the 
phase boundary curve to the small baryochemical potentials
that the {\small QGP} break-up temperature 
$T_f^{\small \rm {\small SPS}}\simeq 145\pm5$ {\small MeV}  
will see just a minor upward change 
to the value $T_f^{\small \rm {\small RHIC}}\simeq 150\pm5$ {\small MeV}.

\begin{figure}[t]
\vspace*{1.5cm}
\centerline{  %\hspace*{-.cm}
\psfig{width=8.5cm,figure=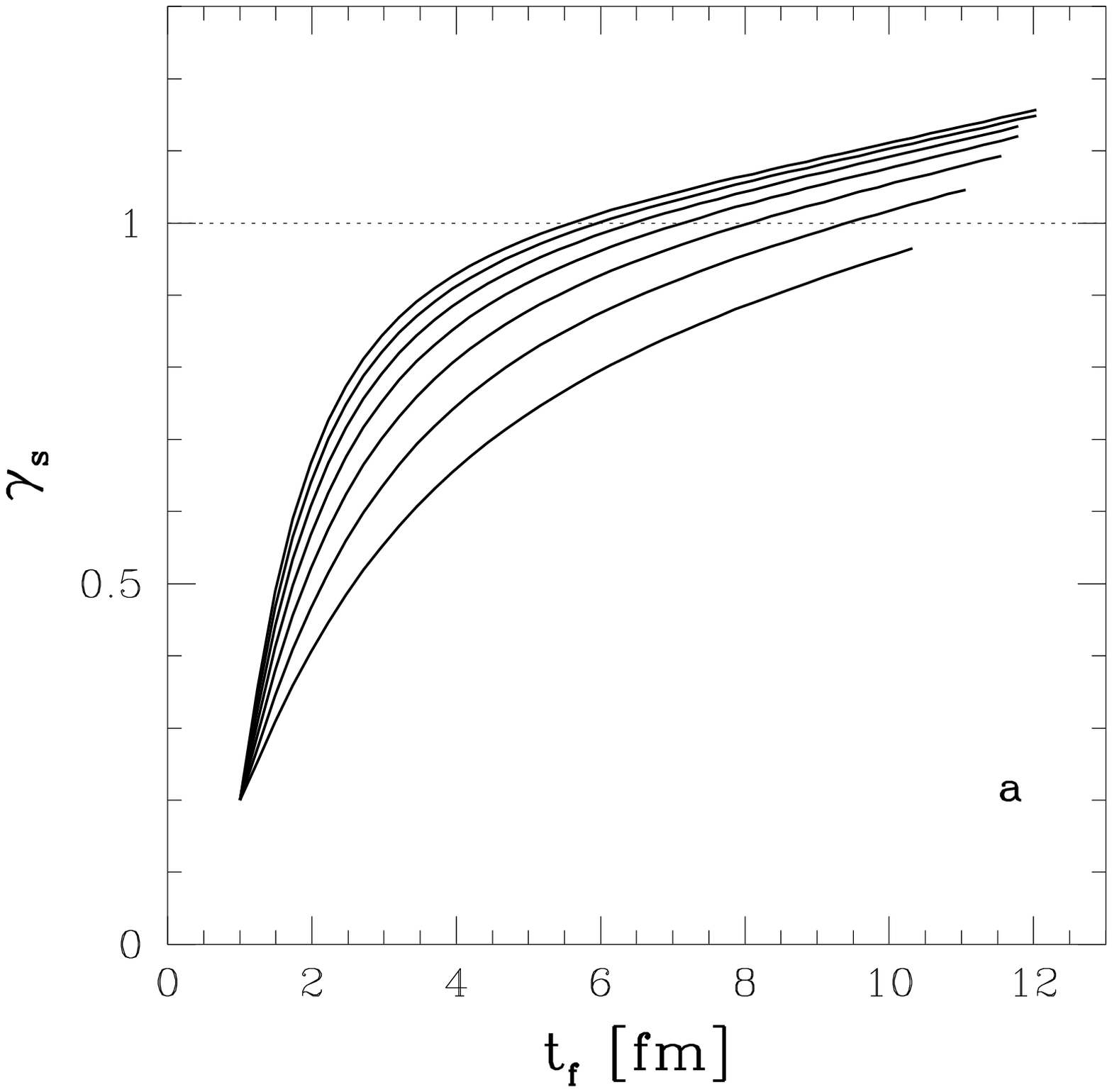}
\vspace*{1.7cm} }\centerline{  %\hspace*{-.cm}   %\hspace*{-.5cm}
\psfig{width=8.5cm,figure=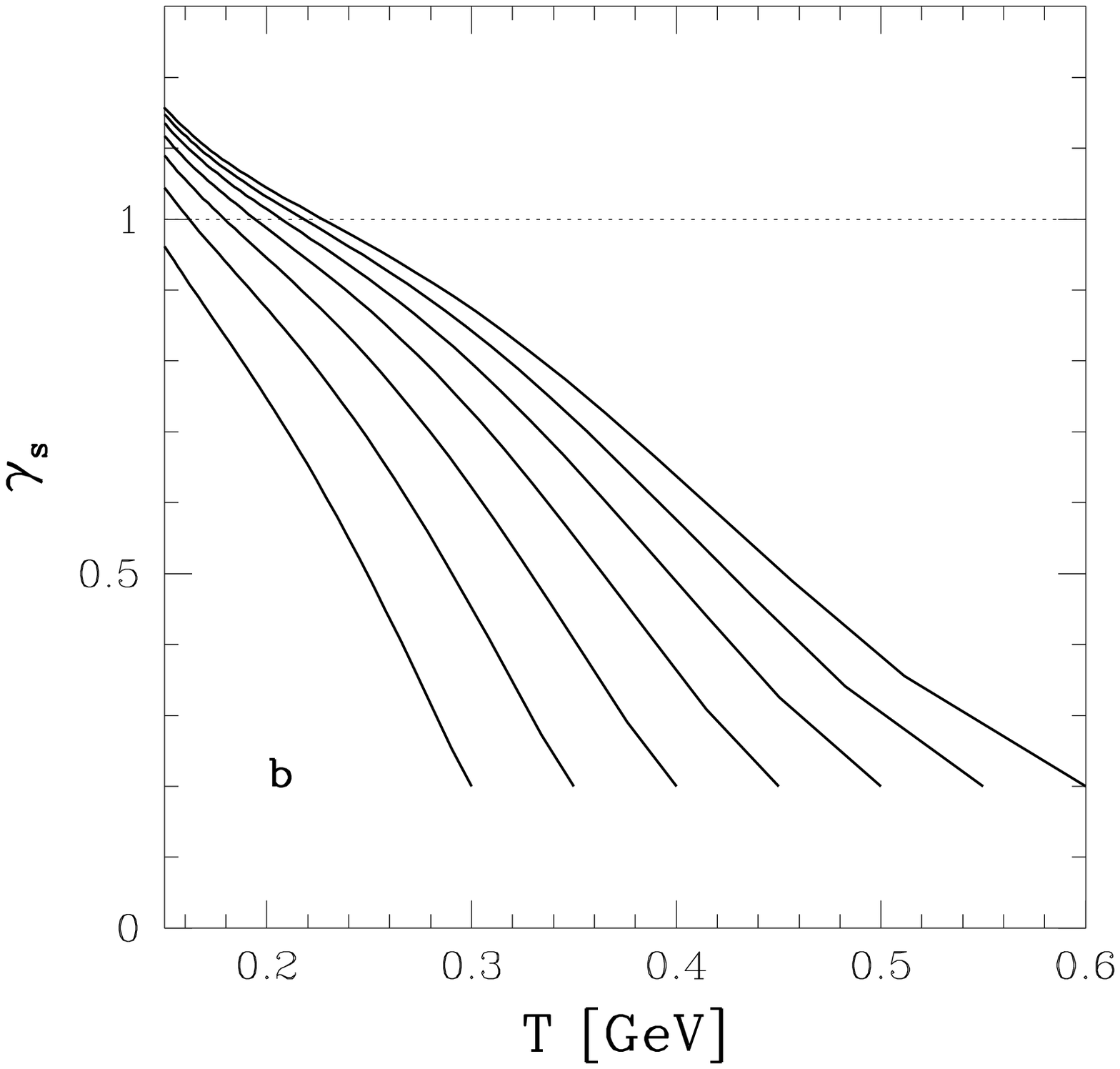}
}
\caption{ %\small
Evolution of QGP-phase strangeness phase space occupancy $\gamma_{\rm s}$ 
{\bf a}) as function of  time and {\bf b}) as function
of temperature, see text for details.
\label{figvarsgam}} 
\end{figure}

With the freeze-out condition fixed, one would think that the major uncertainty 
in our approach comes from the initial gluon equilibration 
temperature $T_0$, and we now study how different values of $T_0$ influence
the final state phase space occupancy. We integrate numerically Eq.\,(\ref{dgdtf}) 
and present $\gamma_{\rm s}$ as function of both time $t$ in
Fig.\,\ref{figvarsgam}a, and temperature $T$ in Fig.\,\ref{figvarsgam}b, 
up to the expected {\small QGP} breakup at 
$T_f^{\small \rm {\small RHIC}}\simeq 150\pm5$ {\small MeV}. 
 We see that: \\
$\bullet$ widely different initial
conditions (with similar initial entropy content) 
lead to rather similar chemical conditions at chemical freeze-out 
of strangeness, \\
$\bullet$  despite
a series of conservative assumptions we find not only that strangeness
equilibrates, but indeed that the dilution effect allows an overpopulation of 
the strange quark phase space.
 For a wide range of initial conditions we obtain a 
narrow band $1.18>\gamma_{\rm s}(T_f)>0.95$\,.  We will in the following 
study of strange  baryon and antibaryon abundances adopt what we believe
to be the  most likely value $\gamma_{\rm s}(T_f)=1.15$.

%%%%%%%%%%%%%%%%%%%%%%%%%%%%%%%
%%%%%%%%%%%%%%%%%%%%%%%%%

We now consider how this relatively large value of $\gamma_{\rm s}$,
characteristic for the underlying {\small QGP} formation and evolution 
of strangeness, impacts the strange baryon and anti-baryon observable
emerging in hadronization.  Remembering that major changes compared to 
{\small SPS} should occur in rapidity 
spectra of mesons, baryons and antibaryons, 
we will apply the same hadronization model that 
worked in the analysis of the {\small SPS} data \cite{recent}:\\
\noindent 1) the {\small QGP}  freeze-out/break-up occurs without a 
significant (transient)  hadronic gas epoch;\\
\noindent 2) the
deconfined {\small QGP} state evaporates over  a few fm/c, during 
which time it remains near to the freeze-out temperature, with 
energy lost due to particle  evaporation and work done against
the vacuum balanced by the internal energy flows. 
This reaction picture can be falsified 
easily, since we expect, based and  compared to the 
Pb--Pb 158$A$ {\small GeV} results:\\
\noindent a) shape identity  
of  {\small RHIC} $m_\bot$ and $y$ spectra of antibaryons 
 $\bar p\,,\ \overline\Lambda\,,\ \overline\Xi\,,$ since 
in our approach there is no difference in 
their production mechanism, and the form of the spectra is determined
in a similar way by the local temperature and flow velocity vector; \\
\noindent b) the $m_\bot$-slopes of these antibaryons
 should be very similar to the result we
have from  Pb--Pb 158$A$ {\small GeV}
since only a slight increase in the freeze-out temperature 
occurs, and no increase in collective transverse flow is expected.

The abundances of particles produced from {\small QGP}  
within this sudden  freeze-out model are controlled by several further chemical 
parameters: the light quark fugacity 
$1<\lambda_q <1.1$\,,  value is limited by the expected small
ratio between baryons and mesons (baryon-poor plasma) when the 
energy per baryon is above 100\,{\small GeV}, strangeness fugacity 
$ \lambda_{\rm s}\simeq 1$ 
which value for locally neutral plasma assures that 
$\langle s-\bar s\rangle =0$; 
the light quark phase space occupancy 
$\gamma_q\simeq 1.5$, overabundance value due to  gluon fragmentation. 
Given these narrow ranges of chemical parameters and 
the freeze-out temperature $T_f=150$ {\small MeV}, 
 we compute the expected particle production at break-up. 
In general we cannot expect that the absolute numbers of particles we
find are correct, as we have not modeled the important effect of flow in
the laboratory frame of reference. However, ratios of
hadrons subject to similar flow effects (compatible hadrons)
can be independent of  the detailed final state dynamics, as 
the results seen at {\small SPS} suggest \cite{recent}, and
we will look at such ratios more closely.
 
Taking $\gamma_q=1.25,\,1.5,\,1.6$ we choose the value of 
$\lambda_q$, see the header  of table \ref{table1}, 
for which the  energy  per baryon ($E/b$)
is similar to the collision condition 
(100\,{\small GeV}),  which leads here to the 
range $\lambda_q=1.03\pm0.005$. We  evaluate for 
these  examples aside of $E/b$, the strangeness per baryon 
$s/b$ and entropy per baryon $S/b$ as shown in the top 
section of the table \ref{table1}. We do
not enforce  $\langle s-\bar s\rangle=0$ exactly, but 
since baryon asymmetry is
small, strangeness is  balanced to better than 2\%\, in
the parameter range considered.
In the bottom portion of  table \ref{table1}  we present
the compatible particle abundance ratios,
computed according to the procedure developed 
in \cite{recent}.  We have presented
aside of the baryon and antibaryon relative yields also the relative 
kaon yield, which is also well determined within our approach.

The meaning of these results can be better appreciated when
we assume in an example the central  rapidity density
of protons is  $dp/dy|_{\mbox{\scriptsize central}}=25$. 
In table  \ref{table2}  we present the 
resulting (anti)baryon abundances. 
We see that the  net baryon density $db/dy\simeq 15\pm2$,  there is 
baryon number transparency. We see that (anti)hyperons are 
indeed more abundant than non-strange  (anti)baryons. It is important
when quoting results from table  \ref{table2} to recall that:\\
\noindent 1) we have chosen arbitrarily the overall
normalization in table  \ref{table2}\,, only particle ratios 
were computed,  and\\ 
\noindent 2) the rapidity baryon density
relation to rapidity proton density is a consequence of the assumed 
value of $\lambda_q$, which we chose to get 
$E/b\simeq 100$\,{\small GeV} per  participant. 

However, we firmly believe that our key result seen in table  \ref{table2}\,, 
the hyperon-dominance of the baryon yields, does not depend on these
hypothesis. Indeed, we have  explored another set of parameters in
our first and preliminary report on this matter \cite{prelim}, finding the same 
primary conclusion explained in the introduction. Another related
notable result, seen in  table \ref{table1}, is that strangeness yield
per participant is 13--23 times greater than seen at present at 
{\small SPS} energies, where we have 0.75 strange quark pairs per baryon.
As seen in  table  \ref{table2} the baryon rapidity density is in our examples
similar to the proton rapidity density.

In summary, we have shown that one can expect strangeness chemical 
equilibration in nuclear collisions at {\small RHIC} if
the deconfined  {\small QGP} is formed, with a probable overpopulation 
effect associated with the early strangeness abundance freeze-out
before hadronization. We studied the physical conditions at {\small QGP}
breakup and have shown that  (anti)hyperons dominate (anti)baryon abundance.

{\it Acknowledgments}: This work was supported in part by a grant from 
the U.S. Department of Energy,  DE-FG03-95OR40937. LPTHE, 
Univ.\,Paris 6 et 7 is: Unit\'e mixte de Recherche du CNRS, UMR7589.
\vskip -.5cm
%%%%%%%%%%%%%%%%%%%%%%%%%%%%%%%%%%%%%%%%%%%%%%%%%%%%%%%%%

%%%%%%%% 
\begin{table}[!t]
\caption{\label{table1} 
For $\gamma_{\rm s}=1.15,\,\lambda_{\rm s}=1$ and $\gamma_q$, $\lambda_q$ as shown:
Top portion: strangeness per baryon $s/b$, 
energy per baryon $E/b$[{\small GeV}]  and  entropy per baryon $S/b$. Bottom portion:
sample of hadron ratios expected at {\small RHIC}.}
\small
\vspace*{-0.2cm}
\begin{center}
\begin{tabular}{llllll}
%\tableline
 $\gamma_q$                               & 1.25 & 1.5  
                      &  1.5  &  1.5  & 1.60 \\
$\lambda_q$                               & 1.03 & 1.025&  1.03 & 1.035 & 1.03 \\
\tableline
$E/b$[{\small GeV}]                       & 117  & 133  &  111  &  96   & 110 \\
$s/b$                                     & 17   & 15   &  12   &  11   & 11 \\
$S/b$                                     & 623  & 693  &  579  & 497   & 567 \\
\tableline
$p/{\bar p}$                              & 1.19 & 1.15 & 1.19  &  1.22 & 1.19 \\
$\Lambda/p$                               & 1.61 & 1.35 & 1.35  &  1.34 & 1.25 \\
${\bar\Lambda}/{\bar p}$                  & 1.71 & 1.41 & 1.42  &  1.43 &1.33 \\
${\bar\Lambda}/{\Lambda}$                 & 0.89 & 0.91 &  0.89 &  0.87 & 0.89 \\
${\Xi}/{\Lambda}$                         & 0.17 & 0.146&  0.145&  0.145& 0.13 \\
${\overline{\Xi}}/{\bar\Lambda}$          & 0.18 & 0.15 &  0.15 &  0.15 & 0.14 \\
${\overline{\Xi}}/{\Xi}$                  & 0.94& 0.95 &  0.94 &  0.93 & 0.94 \\
${\Omega}/{\Xi}$                                    & 0.135&0.114 &  0.113&  0.112& 0.106 \\
${\overline{\Omega}}/{\overline{\Xi}}$              & 0.144& 0.119&  0.120&  0.121& 0.113 \\
${\overline{\Omega}}/{\Omega}$                      &  1   & 1.   &  1.   &  1.   & 1.   \\
$(\Omega+\overline{\Omega})\over(\Xi+\bar{\Xi})$    & 0.14 & 0.12 &  0.12 &  0.12 & 0.11 \\
$(\Xi+\bar{\Xi})\over(\Lambda+\bar{\Lambda})$       & 0.18 & 0.15 &  0.15 &  0.15 & 0.14 \\
${K^+}/{K^-}$                                       & 1.05 & 1.04 &  1.05 &  1.06 & 1.05 \\
%\tableline
\end{tabular}
\end{center}
\vspace*{-.6cm} 
\end{table}
%%%%%%%%%%%%%%%%%%%%

%%%%%%%%%%%%%%%%%%%%
\begin{table}[htb]
\caption{\label{table2} $dN/dy|_{\mbox{\scriptsize central}}$ 
assuming in this example $dp/dy|_{\mbox{\scriptsize central}}=25$ .}
\vspace*{-.2cm} 
\begin{center}
\begin{tabular}{ll|cccccccccc}
$\gamma_q$& $\lambda_q$  & $b$ &  $p$ & $\bar p$ & $\Lambda\!\!+\!\!\Sigma^0$ & $\overline{\Lambda}\!\!+\!\!\overline{\Sigma}^0$&$\Sigma^{\pm}$&$\overline{\Sigma}^{\pm}$ & $\Xi$  
& $\overline\Xi$ & $\Omega\!=\!\overline\Omega$ \\
\tableline
1.25& 1.03 & 16 & 25$^*$& 21 & 40 & 36 & 28 & 25 & 14 & 13 & 0.9  \\ 
1.5 & 1.025 & 13 & 25$^*$& 22 & 34 & 31 & 24 & 21 & 11 & 9.4 & 0.6 \\ 
1.5& 1.03 & 15 & 25$^*$& 21 & 34 & 30 & 24 & 21 & 9.8 & 9.2 & 0.6 \\ 
1.5 & 1.035 & 17 & 25$^*$& 21 & 33 & 29 & 23 & 20 & 9.6 & 9.0 & 0.5 \\ 
1.60& 1.03  & 14 & 25$^*$& 21 & 31 & 28 &22 &19 & 8.6 & 8.0 & 0.5 \\ 
%\tableline
\end{tabular}
\end{center}
%\vspace*{-1.2cm} 
\end{table}

%%%%%%%%%%%%%%%%%%%%

\end{document}